\documentclass[12pt,preprint]{emulateapj}

\newcommand{\msun}{$M_\sun$}

\shorttitle{Star formation rates and stellar masses to $z=7-8$}
\shortauthors{I. Labb\'e et al.}

\def\figsed{
\begin{figure*}
\centering
$$ $$\includegraphics[width=16.75cm,bb=40 360 404 420]{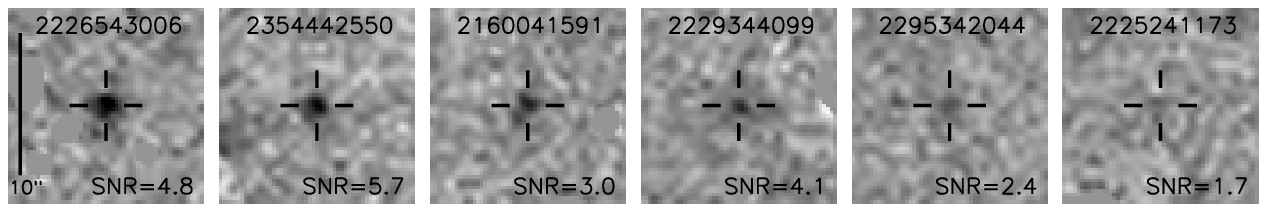}
\makebox[18cm][c]{ \includegraphics[width=9.5cm]{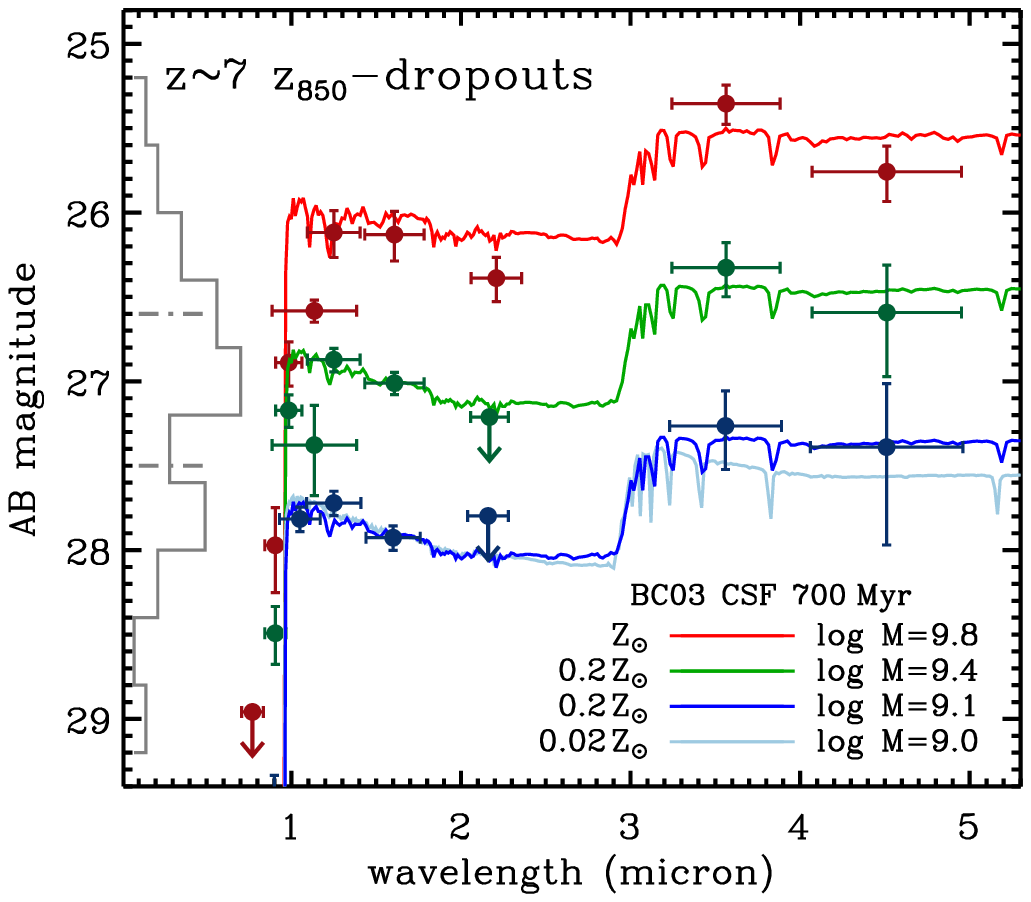}
\includegraphics[width=9.15cm,bb=22 5 362 302]{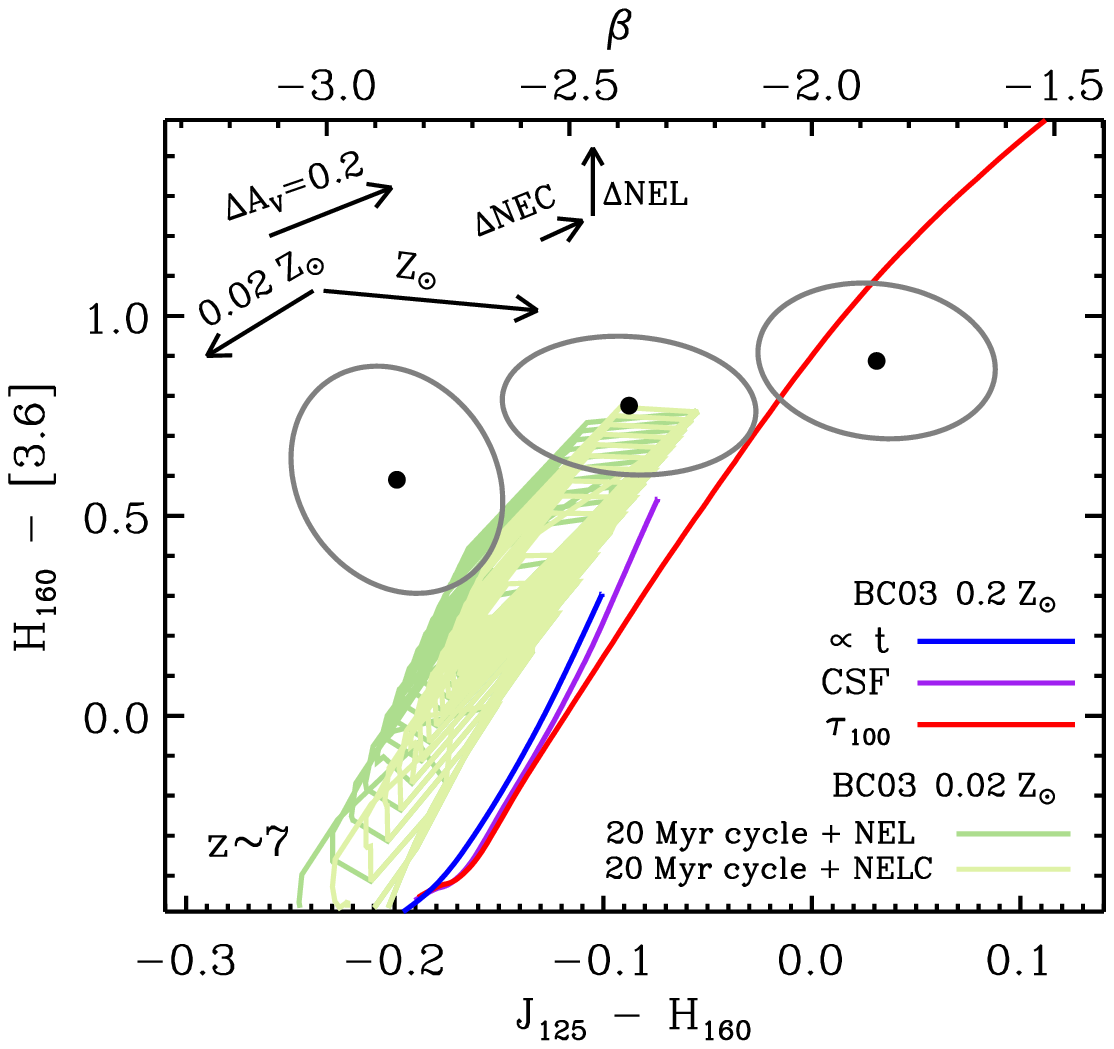} } $$ $$
\leavevmode
\vspace{-1.cm}
\caption{({\it Top panels}) Representative IRAC $[3.6]$ image stamps of 
WFC3/ERS $z\sim7$ $z_{850}-$dropouts (sorted by magnitude) after
subtracting neighbors. Brighter sources may have lower SNR after accounting for confusion.
({\it left panel})  Stacked broadband SEDs from the combined NICMOS, WFC3/UDF and WFC3/ERS samples, 
averaged in $\sim1-$mag bins centered on $H_{160}\approx26,27$~and $28$. 
The data include HST/ACS, NICMOS, and WFC3/IR, groundbased $K$, and IRAC $[3.6]$ and $[4.5]$.
Upper limits are $2\sigma$. 
The gray histogram shows the $H_{160}$-band magnitude distribution (the peak is 9 sources).
The best-fit BC03 stellar population models at $z=6.9$ are shown. The overall SED shapes are 
similar with a break between $H_{160}$ and $[3.6]$, expected for evolved stellar populations ($>100$Myr). 
The far$-UV$ slope (traced by $J_{125}-H_{160}$) bluens towards fainter $H_{160}$ magnitude 
(as found \citealt{Bo09b}).
({\it right panel}) Comparison of the observed average 
$J_{125}-H_{160}$ versus $H_{160}-[3.6]$ color ({\it filled circles}) 
with predictions of $0.2~Z_\sun$ BC03 models for various SFHs ({\it solid lines}). 
Ellipsoids show $1\sigma$ uncertainties.
The SFHs are linearly increasing with time ({\it blue}), CSF ({\it purple}), 
episodic ({\it light green}) with a 50\% dutycyle and 40Myr duration, and 
exponentially declining ({\it red}) with $\tau=100$Myr. 
Also shown are the effects on the model colors of metallicity, reddening 
(Calzetti et al. 2000), nebular continuum emission (NEC) \citep{Sc02}, 
and line emission (NEL) \citep{Erb06,Bri08}. The episodic model is shown twice, once 
only with the effect of line emission (+ NEL) and once with both 
line and nebular continuum emission (+ NELC). The lines terminate at the 750Myr.
\label{figsed}}
\end{figure*}
}

\def\figsfr{
\begin{figure}
\epsscale{1.2}
\centering
$$ $$\includegraphics[width=6.7cm,bb=43 390 425 527]{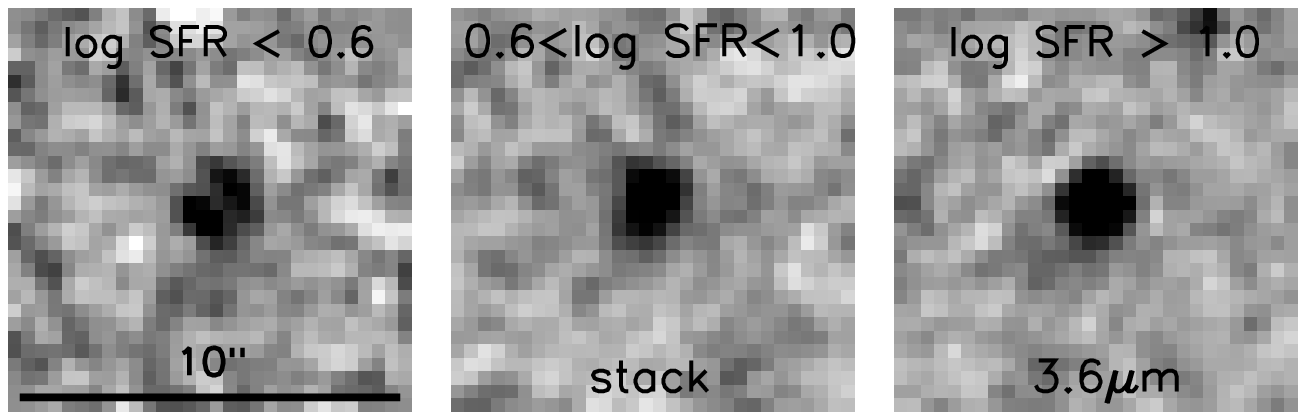}
\includegraphics[width=9.cm]{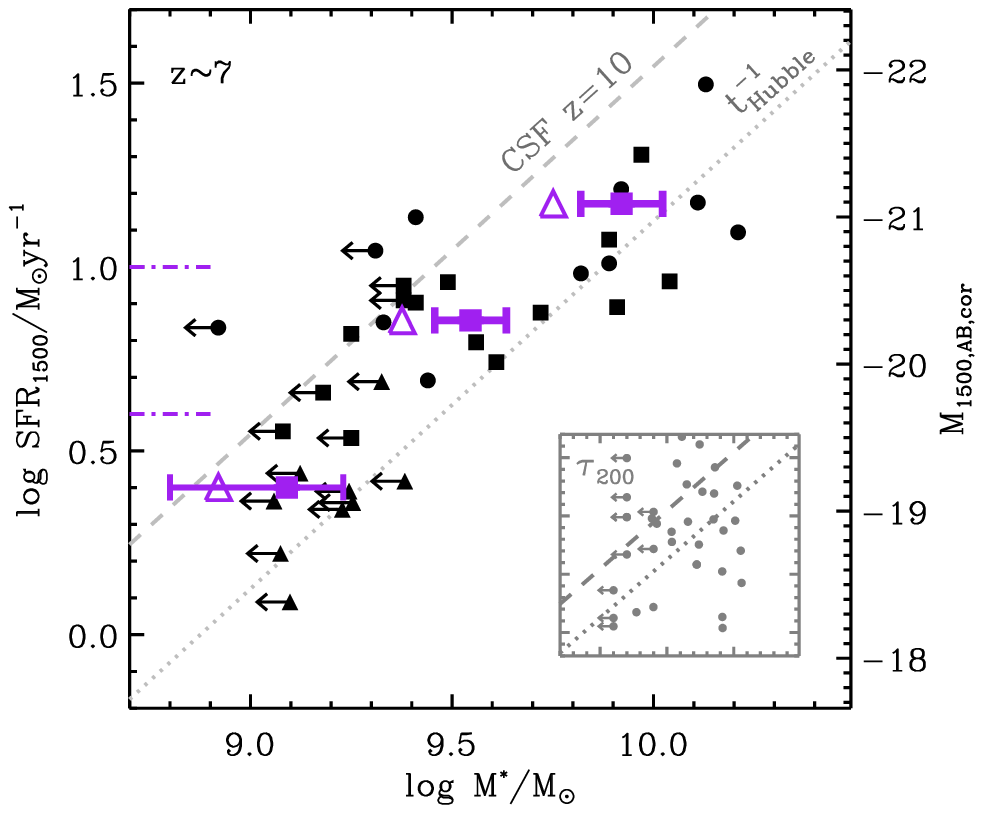}$$ $$
\leavevmode
\vspace{-1.2cm}
\caption{Average $UV-$derived, dust-corrected SFR versus stellar mass for $z\sim7$ galaxies.
Black symbols denote galaxies from the NICMOS ({\it circles}), WFC3/IR ERS ({\it squares}) 
and WFC3/IR HUDF sample ({\it trangles}). 
The purple squares show the average stellar mass in bins of SFR centered 
on $\log SFR \approx 0.4, 0.8, 1.2$. The purple triangles 
assume $\approx0.2$ and $0.15$ mag contribution by emission lines to the $[3.6]$
and $[4.5]$ bands. The scatter in $M^*$ in bins of $M_{1500}$ is $\sim0.3~$dex. 
The diagonal lines show the maximum stellar mass CSF stellar population
can form in a Hubble time ({\it dotted}) or since 
$z=10$ ({\it dashed}). Galaxies with strongly increasing/declining SFRs 
would lie well above/below the lines, respectively, but few such systems are found. 
The inset shows simulated galaxies with random formation times and exponentially 
declining $\tau=200~$Myr SFHs, showing a different distribution with larger scatter.
($Top\ panels$) Stacked images in the [3.6] band in bins of SFR.
\label{figmsfr}}
\end{figure}  
}

\def\figy{
\begin{figure}
\centering
$$ $$\includegraphics[width=6.98cm,bb=-62 38 504 415]{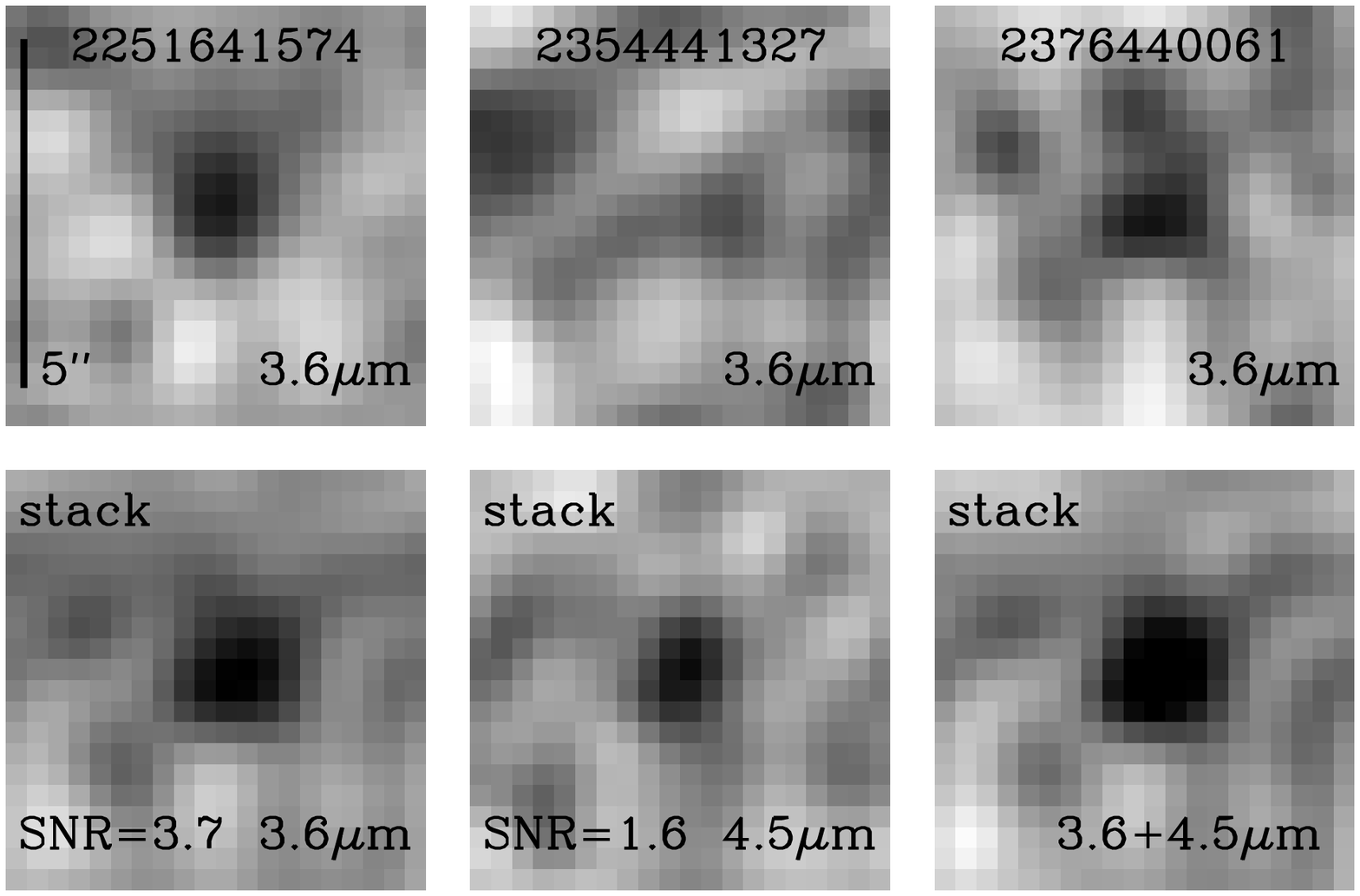}
\includegraphics[width=9.cm]{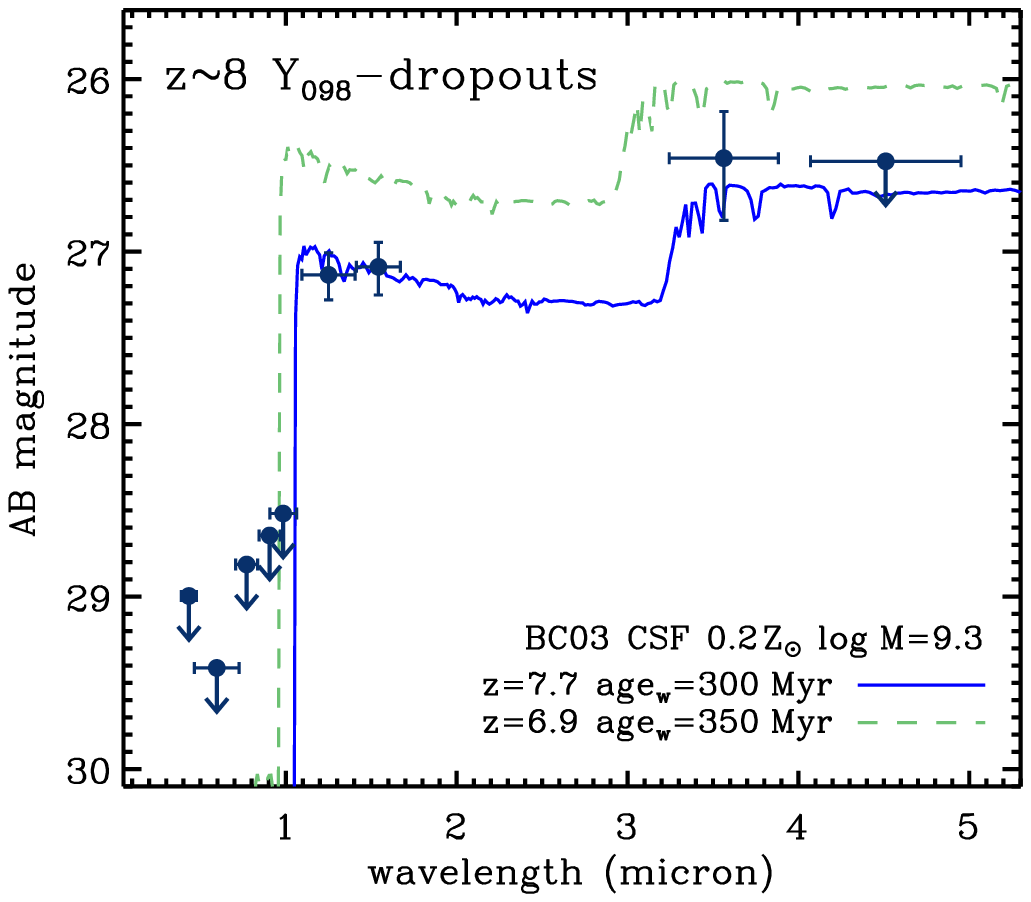}$$ $$
\vspace{-1cm}
\caption{
({\it top panels}) Individual and stacked images of the 3 $z\sim8$ $Y_{098}-$dropout 
galaxies in our WFC3/ERS sample, shown in inverted grayscale. 
Two of the 3 sources are individually detected in the IRAC $[3.6]-$band. 
({\it bottom panel}) The average broadband
SED of the 3  $z\sim8$ galaxies and the best-fit stellar population model 
({\it solid blue line}). For comparison, we show the best-fit model to $z=6.9$ 
$z_{850}$ dropouts of similar $H_{160}$ magnitude ({\it dashed green line}), 
shifted by $-0.5$ mag. The overall shapes are similar, with the $z\sim8$ galaxies
being slightly bluer in $H_{160}-[3.6]$ compared to $z\sim7$ galaxies. 
Upper limits are $2\sigma$. 
\label{figydrop}}
\end{figure}
}

\def\figmass{
\begin{figure}
\epsscale{1.23}
\centering
\plotone{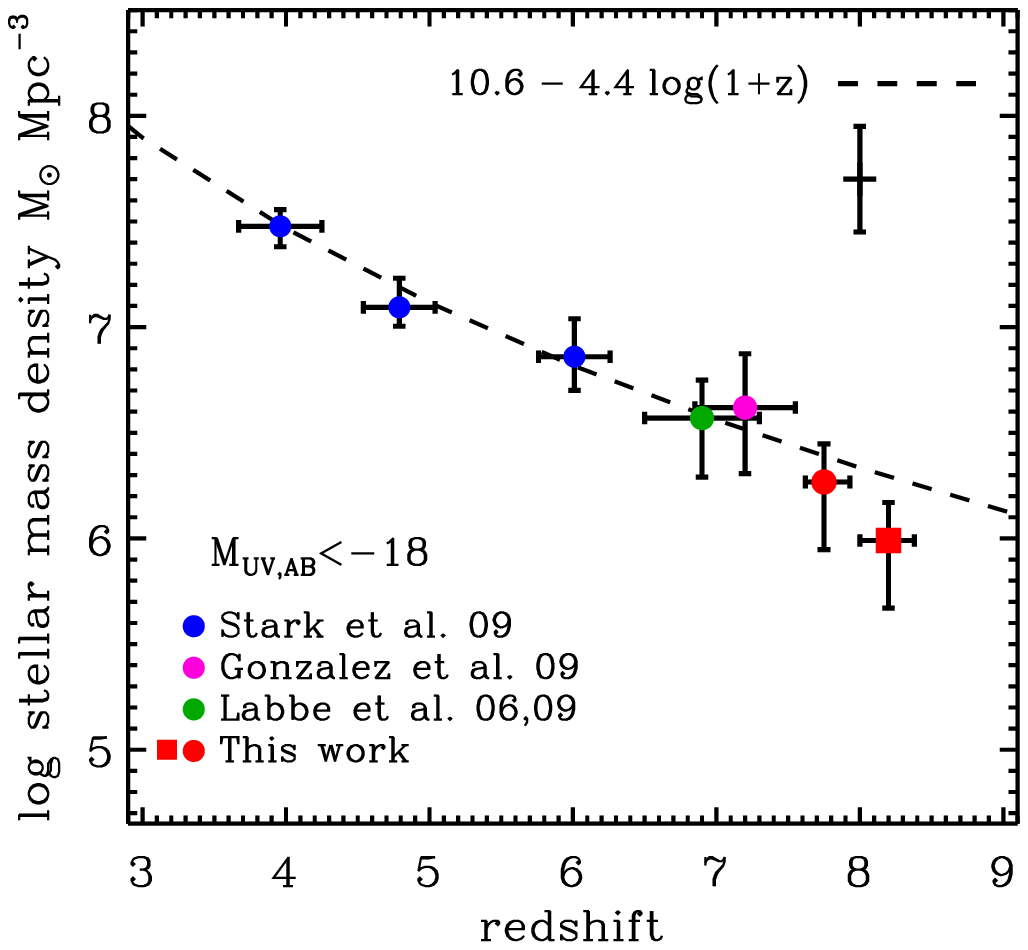}
\caption{The evolution of the integrated stellar mass density. 
The red circle shows the $z\approx7.7$ mass density, derived from
the integrated $UV-$luminosity density of (Bouwens
et al., in preparation) and the mean $M/L$ derived here.
All data are converted to a common limit $M_{UV,AB}<-18$,
using the $UV$ LFs of \citet{Bo09a}. The WFC3/ERS data 
are corrected by +0.42 dex. The $z=3-7$ luminous samples ($M_{UV,AB}<-20$) 
from the literature (\citealt{St09} {\it blue 
circles}, \citet{Go09}, {\it magenta circle}) 
are corrected by +0.38, +0.46, +0.57, and +0.75 dex at z=4, 5, 6, and 7, respectively.
The HUDF requires no correction (\citealt{La09}, {\it green circle}).
The red square shows the $z\approx8.2$  sample of 
\citep{Bo09a}, but using the $M/L$ derived here for the
$z\approx7.7$ sources. The dashed line shows a $\propto(1+z)^{-4.4}$ evolution.
The floating error bar indicates the expected cosmic 
variance for the $z\sim8$ sample. 
\label{figmass}}
\end{figure}
}

\begin{document}

\title{Star formations rates and stellar masses of $z=7-8$ galaxies \\
from IRAC observations of the WFC3/IR ERS and the HUDF fields\altaffilmark{1}}

\author{I. Labb\'e\altaffilmark{2,3} 
V. Gonz\'alez\altaffilmark{4}, 
R. J. Bouwens\altaffilmark{4,5}, 
G. D. Illingworth\altaffilmark{4}, 
M. Franx\altaffilmark{5}, 
M. Trenti\altaffilmark{6}, 
P. A. Oesch\altaffilmark{7},
P. G. van Dokkum\altaffilmark{8}
M. Stiavelli\altaffilmark{9},
C. M. Carollo\altaffilmark{7}, 
M. Kriek\altaffilmark{10},
D. Magee\altaffilmark{4}
}

\altaffiltext{1}{Based on observations made with the NASA/ESA Hubble Space
Telescope, which is operated by the Association of Universities for
Research in Astronomy, Inc., under NASA contract NAS 5-26555. These
observations are associated with programs \#11563, 9797. Based on observations 
with the {\em Spitzer Space Telescope}, which is operated by the Jet 
Propulsion Laboratory, California Institute of Technology under 
NASA contract 1407.  Support for this work was provided by NASA through contract 
125790 issued by JPL/Caltech. Based on service mode observations collected at 
the European Southern Observatory, Paranal, Chile (ESO Program 073.A-0764A).
Based on data gathered with the 6.5 meter Magellan Telescopes located at Las Campanas 
Observatory, Chile.}
\altaffiltext{2}{Carnegie Observatories, Pasadena, CA 91101}
\email{ivo@obs.carnegiescience.edu}
\altaffiltext{3}{Hubble Fellow}
\altaffiltext{4}{UCO/Lick Observatory, University of California, Santa Cruz, CA 95064}
\altaffiltext{5}{Leiden Observatory, Leiden University, NL-2300 RA Leiden, Netherlands}
\altaffiltext{6}{University of Colorado, Center for Astrophysics and Space Astronomy, 389-UCB, Boulder, CO 80309, USA}
\altaffiltext{7}{Institute for Astronomy, ETH Zurich, 8092 Zurich, Switzerland}
\altaffiltext{8}{Department of Astronomy, Yale University, New Haven, CT 06520}
\altaffiltext{9}{Space Telescope Science Institute, Baltimore, MD 21218, United States}
\altaffiltext{10}{Department of Astrophysical Sciences, Princeton University,Princeton, NJ 08544}

\begin{abstract}
We investigate the Spitzer/IRAC properties of 36 $z\sim7$ $z_{850}-$dropout
galaxies and 3 $z\sim8$ $Y_{098}$ galaxies derived from deep/wide-area 
WFC3/IR data of the Early Release Science, the ultradeep HUDF09, and wide-area NICMOS data. 
We fit stellar population synthesis models to the SEDs to derive mean
redshifts, stellar masses, and ages. The $z\sim7$ galaxies
are best characterized by substantial ages ($>100$~Myr) and $M/L_V\approx0.2$. The main trend 
with decreasing luminosity is that of bluing of the far$-UV$ slope from $\beta\sim-2.0$ 
to $\beta\sim-3.0$. This can be explained by decreasing metallicity, except for the 
lowest luminosity galaxies ($0.1~L^*_{z=3}$), where low metallicity and smooth SFHs
fail to match the blue far$-UV$ and moderately red $H-[3.6]$ color.
Such colors may require episodic SFHs with short periods of activity and
quiescence (``on-off'' cycles) and/or a contribution from emission lines. 
The stellar mass of our sample of $z\sim7$ star forming galaxies 
correlates with SFR according to $\log M^{*}=8.70(\pm0.09)+1.06(\pm0.10)\log SFR$,
implying star formation may have commenced at $z>10$.
No galaxies are found with SFRs much higher or lower than the past 
averaged SFR suggesting that the typical star formation timescales are probably
a substantial fraction of the Hubble time.
We report the first IRAC detection of $Y_{098}-$dropout galaxies at $z\sim8$. 
The average rest-frame $U-V\approx0.3$ (AB) of the 3 galaxies are similar to faint 
$z\sim7$ galaxies, implying similar $M/L$. The stellar mass
density to $M_{UV,AB}<-18$ is $\rho^*(z=8)=1.8^{+0.7}_{-1.0}\times10^6~M_\sun~$Mpc$^{-3}$,
following $\log\rho^*(z)=10.6(\pm0.6) - 4.4(\pm0.7) \log(1+z)$ $[M_\sun~$Mpc$^{-3}]$ over $3<z< 8$.
\end{abstract}

\section{Introduction}
Until recently, only a modest number of relatively bright
$z\gtrsim7$ galaxies were known, mostly from wide-area NICMOS 
\citep[][R.~J.~Bouwens et al. in preparation]{Bo08,Oe09a} 
and ground-based searches\citep{Ou09,Ca09,Hi09}. 
The arrival of WFC3/IR aboard HST has dramatically improved 
the situation by identifying large numbers of 
$z\gtrsim7$ galaxies by their redshifted $UV$ light.
Here we report on $z\gtrsim7$ galaxies selected 
from the WFC3/IR Early Release Science (ERS) observations 
over the GOODS-South field (\citealt{Wil09}, R.~J.~Bouwens et al. in preparation),
complemented with candidates from the recent ultradeep survey with 
WFC3/IR over the HUDF09 field (\citealt{Oe09b,Bo09a}; see also \citealt{Mc09,Bu09,Ya09}).

Little is known about the stellar masses, metal production, and the 
contribution of star formation to reionization in these galaxies.
Mid-infrared observations with the InfraRed 
Array Camera (IRAC; \citealt{Fa04}) on {\it Spitzer} 
can been used to constrain the stellar masses and ages, which has 
lead to the surprising discovery of quite massive $\sim10^{10}M_\sun$ 
galaxies at $z\gtrsim6$ \citep{Ey05,Ya06,St09} and appreciable ages
($200-300$Myr) and $M/L$s as early as $z\sim7$ 
\citep{Eg05,La06,Go09}. The overall results suggest the galaxies formed 
substantial amounts of stars at even earlier times, well into 
the epoch of reionization \citep{St07, Ya06, La09}.

In this Letter, we study the stellar populations of the largest 
sample of $z\gtrsim7$ galaxies with IRAC measurements to date,
focusing on correlations with luminosity and stellar mass, 
and implications for the mass density and $z\sim8$.
We adopt an $\Omega_M=0.3,\Omega_\Lambda=0.7$ cosmology with 
$H_0=70$~km~s$^{-1}$Mpc$^{-1}$. Magnitudes are in the AB 
photometric system \citep{Ok83}. 

\section{Observations and Stellar Population Modeling}

Our sample consists of sources derived
from the ultradeep WFC3/IR HUDF, the deep WFC3/IR ERS,
and wide-area NICMOS over the CDF-S and CDF-N. Candidates were
selected using the $z\sim7$ $z_{850}-$dropouts and $z\sim8$ $Y_{098}-$dropouts,
as used in \citet{Oe09b}, \citet{Bo09b}, and \citet{Go09}
(see also R.~J.~Bouwens et al. in preparation). We now 
briefly discuss the IRAC photometry from the new ERS sample. 

The Spitzer/IRAC data over the WFC3/ERS 
area in the CDF-S ($\approx23.3$ hours integration time) was obtained from the
Great Observatories Origins Deep Survey (GOODS; M. Dickinson et al. in preparation).
\footnote{This paper uses data release DR2 of epoch 2, 
available from \url{http://data.spitzer.caltech.edu/popular/goods/}}. 
The IRAC depths in the $3.6$ and $4.5-$bands are $27.1$ and $26.5$ magnitude
($1\sigma$, total, point source), respectively. 
Obtaining reliable IRAC fluxes of the candidates is challenging 
because of the contamination from the extended PSF wings of 
nearby foreground sources. 
We remove contaminating flux, by modeling
the candidates and nearby sources using their isolated 
flux profiles and positions in the deep WFC3/IR maps as 
templates. We convolve the templates to match the IRAC PSF, 
simultaneously fit them to the IRAC map leaving only the
fluxes as free parameters, and subtract the best-fit models to
the foreground sources (see \citealt{La06,Wu07,Go09,dSa07}).
After cleaning the IRAC images, we perform conventional 
aperture photometry in the $3.6$ and $4.5$ bands
in 2\farcs5 diameter apertures on 15 of the original 
18 $z_{850}-$dropout galaxies over the WFC3/IR ERS. Three 
were too close to bright sources for 
reliable measurement. Fluxes were corrected by a factor
$\times1.8$ to account for light outside the aperture 
(consistent with point source profiles). 
Six of the 15 galaxies are undetected in IRAC 
($[3.6]<26.5$,$2\sigma$). Errors include the uncertainty
in the best-fit confusion correction, added in quadrature.

We complete the sample with 21 $z\sim7$ 
galaxies with IRAC measurements from the HUDF 
\citep{Oe09b,La09} and from the recent wide-area NICMOS
search \cite[][R.~J.~Bouwens, in preparation]{Go09}. The
total sample consists of 36 $z\sim7$ galaxies spanning
4 magnitudes in $H_{160}$. To investigate trends with 
magnitude, we stacked the flux densities of the galaxies 
in three $\sim1-$magnitude bins centered on 
$H_{160}\approx26,27$ and $28$, containing 11, 15, and 
10 galaxies respectively. The uncertainties are determined 
by bootstrapping. Stacking increases the SNR, in
particular for faint galaxies, where the uncertainty
in the mass is driven by the SNR in IRAC (see
\citealt{La10}).

We derive stellar masses and redshifts by fitting stellar populations 
synthesis models to the average SED fluxes using 
the $\chi^2-$fitting code FAST \citep{Kr09}.
We adopt \citet[][BC03]{bc03} models with a \citet{Sa55} initial 
mass function (IMF) between $0.1-100$~\msun. 
We explore several SFHs and the effects of metallicity and dust.
The differences with more recent models (e.g., \citealt{Ma05},
Charlot \& Bruzual, in preparation) are small and will not be 
considered in detail (see \citealt{La09}). 
We fit models smoothed to a resolution 
of 100~\AA\ rest-frame, corresponding to the approximate width of the dropout 
selection windows. Adopting a \citet{Kr01} IMF
reduces the stellar masses and SFRs by $0.2$~dex, but does 
not change other parameters or the quality of fit. 
The typical uncertainties in the derived average stellar masses, 
SFRs, age, and $A_V$ for the stacked SEDs are 0.15~dex, 0.25~dex, 0.3~dex, and 
0.1~mag, respectively. The photometry and best-fit model parameters 
are presented in Table~1.

\section{Stellar populations and Star Formation Histories at $z\sim7$}
\label{sec.SFH}
Figure~1 ({\it left panel}) shows the broadband SEDs 
of the $z\sim7$ $z_{850}-$dropout
galaxies in the three magnitude bins, with the best-fit BC03 
stellar population models. The overall SED shapes are remarkably similar,
with a pronounced jump between $H_{160}$ and $[3.6]$ (or rest-frame 
$(U-V)\approx0.5$) indicative of a modest Balmer break and 
 evolved stellar populations ($>100$Myr). Focusing on the 
far$-UV$ continuum, we find the slope $f_\lambda\propto\lambda^\beta$ 
(traced by the $J_{125}-H_{160}$ color) to be very blue, and
decreasing from $\beta\sim-2$ at $H_{160}\sim26$ to $\beta\sim-3.0$ 
at $H_{160}\sim28$. As discussed by \citet{Bo09b}, such extremely blue slopes
require low dust content $A_V<0.1$, very low metallicities 
and/or very young ages. Small $0.04$ mag changes in the WFC3/IR zeropoints 
(e.g., \cite{Mc09}) would cause changes of $\beta\approx0.17$, 
comparable to the random uncertainties.
The red $H_{160}-[3.6]$ color, however, implies 
more evolved stellar populations ($>100$Myr), leaving the models seemingly 
unable to match the entire SED.

To explore the mismatch further we consider in more detail 
the effects on metallicity, SFH, and nebular emission. 
Fig.~1 ({\it right panel}) shows the $J_{125}-H_{160}$ versus 
$H_{160}-[3.6]$ colors of the observed stacked SEDs. 
The lines show predictions of $0.2~Z_\sun$ BC03 models for various
continuous SFHs (rising, constant, declining). Generally, evolved
models are able to reproduce the joint $J_{125}-H_{160}$ versus $H_{160}-[3.6]$ 
colors of the more luminous $z=7$ galaxies, but not the colors of the 
faintest, bluest galaxies. The arrows show the effect of changes in
model assumptions, which we will discuss now:

\figsed

{\it 1) Metallicities:} Low metallicities (e.g., $0.2~Z_\sun$)  do 
a decent job of producing much bluer $\beta$ than Solar at a given 
$H_{160}-[3.6]$ color. Very low metallicities ($1/50$ Solar) produce 
even bluer $\beta$'s, but also bluen $H_{160}-3.6$. 
Metallicity alone appears not enough to fully resolve the discrepancy.

{\it 2) Nebular Emission:} 
Nebular emission lines (NEL) likely contribute to 
the IRAC fluxes, reddening the $H_{160}-[3.6]$ color.
Empirical estimates of [OIII]5007 emission at $z>2$ are scarce, but we can 
infer the possible effect from the observed strength of H$\alpha$ at $z\sim2.2$ 
\citep{Erb06}. Assuming W$_{H\alpha}=200\AA$ and W$_{[OIII]4959,5007+H\beta}= 2.5 \times W_{H\alpha}$,
appropriate for $0.2~Z_\sun$ galaxies at $z>2$ \citep{Erb06,Bri08}, and
adopting a redshift distribution of $z=6.9\pm0.5$ \citep{Oe09b} we 
calculate a contribution of 0.18 mag to $[3.6]$ and 0.14 mag at $[4.5]$
(see Fig~1).  Note however, that no measurements of nebular lines at $
z\sim7$ exist. A second possible effect is nebular 
continuum emission (NEC), which would cause a reddening of 
both $\beta$ and $H_{160}-[3.6]$ (shown are the models of \citealt{Sc02} 
with $Z=1/50~Z_\sun$, $t=300~$Myr CSF, and 0\% escape fraction).
Both would reduce the discrepancy between models and data by allowing the models 
to extend to redder $H_{160}-[3.6]$ at a given $\beta$. 
Note that the extremely blue $\beta$ may require high escape 
fractions $f_{esc}\gtrsim0.3$ (e.g., \citealt{Bo09b}), which would
reduce the contribution of nebular emission.

{\it 3) Star Formation Histories:} 
Declining SFHs can match the colors of the most luminous, redder galaxies, but their
far$-UV$  continua are too red for lower luminosity galaxies.
In contrast, strongly rising SFRs 
(SFR$\propto~t^\alpha$, $\alpha>1$) exhibit blue $\beta$ but never reach 
red $H_{160}-[3.6]$ in a Hubble time $(z=7)$ and are formally excluded 
at 95\% confidence. CSF is an compromise, providing red 
$H_{160}-[3.6]$ colors generated by on-going assembly 
of stellar mass, and  blue far$-UV$ continua from on-going star formation.
Finally, episodic SFH with a 50\% duty cyle and 
40~Myr duration (i.e., 20~Myr ``on'', 20~Myr ``off'') are found to have
an interesting mix of properties.  The luminous active phase of 
the cycle produces a bluer far$-UV$ continuum for a given $H_{160}-[3.6]$ color 
than CSF. The reverse is true in the dimmer passive state. The 
net result for a steep $UV$ LF function is that the luminosity weighted 
average of cycling galaxies displays bluer far$-UV$ at a 
given $H_{160}-3.6$ than CSF, also reducing the discrepancy.

In summary, the model colors match the observations of luminous $z\sim7$
galaxies reasonably except for the lowest luminosity galaxies, where
the blue $\beta\sim-3$ and the red $H_{160}-[3.6]\sim0.6$ colors remain
challenging to fit. Low metallicity CSF models 
come close, but a contribution from nebular line emission to the $[3.6]-$band
and/or episodic SFHs are likely needed to resolve the mismatch. 

\figsfr
\figy

\section{Star Formation Rate versus Stellar Mass at $z\sim7$}

Independent constraints on the SFHs can be obtained from the relation 
between SFR and stellar mass, as shown in Fig.~2. 
The SFRs are calculated from the monochromatic $1500\AA$
luminosity following the prescription of \citet{Ma98} and 
corrected for dust using the best-fit $A_V$. The 
galaxies are grouped in bins of SFR (or $M_{1500}$) centered on 
$\log SFR\approx0.4, 0.8, 1.2$. 

The stellar mass of our $z\sim7$ sample correlates 
strongly with the SFR, producing  $\log M^*=8.70(\pm0.09)+1.06(\pm0.10)\log~SFR$
and corresponding to a constant $M/L_{V}\approx0.20$.\footnote{
The M/L also be calculated from the individual galaxies, yielding the same
answer \citep{Go09}.}
The derived stellar mass with emission lines (see \S3) would be 
lower by $\approx0.17$dex. The scatter around the relation is fairly 
low $\approx0.3~$dex, but
that does not exclude significant short-term SFR variation, 
e.g., the episodic model (\S\ref{sec.SFH}) predicts 
$\log M^*=8.45+1.0\log~SFR_{1500}$ and a scatter of $\sim0.3$~dex.
We find no galaxies with SFRs much lower or higher than the past 
averaged SFR (i.e., strongly bursting or suppressed. Such 
galaxies would have satisfied our dropout criteria and would lie
in the upper left or lower right corner in Fig.~2. Their absence
suggests that that the typical star formation timescales are probably 
a substantial fraction of the Hubble time.
Instead we find that only  4/22 sources with $\log SFR>0.7$ are undetected at $[3.6]$ and no galaxies 
in the sample have SFRs substantially less than $M^*/t_{Hubble}$. 

To illustrate the diagnostic power of the $M^*-SFR$ diagram, we show in the 
inset in  Fig.~2 a simulation of galaxies with random formation times 
and exponentially declining SFRs ($\tau=200~$Myr) to the same selection 
limits as our observed sample. The distribution is clearly different, 
with no correlation between $M^*$ and SFR, suggesting that star 
formation timescales for $z\sim7$ galaxies are probably longer than that.

\figmass

\section{The stellar mass density at $z\sim8$}

The detection of $z\sim8$ galaxies with IRAC is enticing as 
it enables us to place stronger constraints on the stellar
masses of the highest redshift galaxies than possible from the far$-UV$ alone.
Recent studies in the HUDF have found no detection
(individual or stacked) for $z\sim8$ galaxy candidates \citep{La09}, 
leaving estimates of the stellar mass density at these redshifts
highly uncertain.

Here we perform photometry on the 3 $z\sim8$ $Y_{098}-$dropout galaxies 
in the WFC3/ERS sample of R.~J. Bouwens et al. (in preparation). These 
candidates are brighter than found in the HUDF 
($H\approx27$ versus $H\approx28$). Two galaxies
are detected at $[3.6]$\footnote{We caution that source ERSy-2376440061 at $[3.6]$ is close to pixels 
that are affected by ``Muxbleed'', which we subtracted using a 3rd order 
polynomial fit to the $20\arcsec\times20\arcsec$ background before performing photometry.} 
and we calculate an average $SNR=3.7$ in $[3.6]$ and $SNR=1.6$ in $[4.5]$ for the stack of 
all three (see Fig~3, top panels). 
The best fits are $z=7.7^{+0.18}_{-0.15}$, high stellar 
age $age_w=300^{+50}_{-210}~$Myr\footnote{Following \cite{La06}
we report {\em SFH weighted} age$_w$, where age$_{w}=t/2$ for CSF and $t$ is the time 
elapsed since the start of star formation.}, mass-to-light ratios $M/L_{V}=0.15$ 
and $M/L_{1500}=0.1$, and $\log SSFR = -8.7$.
Overall these properties are comparable to $H\approx27$ $z\sim7$ $z_{850}-$dropouts, 
suggesting modest evolution in the $M/L$ between $z=8$ and $z=7$.

Following the approach of \citet{Go09} and \citet{La09}, we derive integrated stellar 
mass densities at $z\sim8$ by multiplying the $UV-$luminosity densities 
integrated to $M_{UV,AB}=-18$ \citep[][R~.J~.Bouwens et al. in preparation]{Bo09a}
by the mean $M/L$ derived for the $z\sim7.7$ galaxies,
yielding $\rho^*(z=8)=1.8^{+0.7}_{-1.0}\times10^6~M_\sun~$Mpc$^{-3}$.
We also recompute the $z\sim8$ stellar mass density of \citet{La09} using
the same $M/L$. Figure~4 shows the evolution of the stellar mass density 
from $z=3$ to $z=8$. The evolution over $3<z<8$ is well approximated by 
$\log\rho^*(z)=10.6(\pm0.6) - 4.4(\pm0.7) \log(1+z)$ $[M_\sun~$Mpc$^{-3}]$ over $3<z< 8$.

\section{Summary}
Using a large sample of 36 $z_{850}-$dropout galaxies based on 
deep/wide-area WFC3/IR data from the Early Release Science, ultradeep data from 
the HUDF09, and wide-area NICMOS programs, we investigate the stellar population 
properties at extreme redshifts $z\gtrsim7$. The main results are:

\begin{itemize}
\item The average rest-frame far$-UV$ slope at $z\sim7$ becomes bluer 
with decreasing luminosity, from $\beta\sim-2.0$ 
($L^*_{z=3}$) to $\beta\sim-3.0$ ($0.1~L^*_{z=3}$), as reported by \cite{Bo09b}.
The rest-frame $U-V$ becomes bluer as well, but is still moderately
red $U-V\approx0.3$ at $0.1~L^*_{z=3}$, apparently excluding extremely young ages 
$<100$Myr. If the ages inferred from the simple model fits are 
correct, galaxies started forming stars very early-on, perhaps as high as 
$z>10$. The blue far$-UV$ slope and red  $U-V$ colors remain a challenge to fit,
however, even for sub-solar metallicity models.
Episodic SFHs with periods of activity and quiescence and/or a ($\sim0.2$ mag) 
contribution of emission lines to the $[3.6]-$band may be required to resolve 
the mismatch.
\item The derived stellar masses correlate with the SFRs at $z\sim7$ according
to $\log M^*=8.70(\pm0.09)+1.06(\pm0.10)\log~SFR$, with relatively low scatter 
$\sim0.25~$dex. Emission line contributions of  $\approx0.2$ mag to both $[3.6]$ and $[4.5]$ 
would shift the relation by $\approx-0.2$ dex in mass. The absence of galaxies with 
SFRs much lower or higher than the past averaged SFR (i.e., strongly bursting or suppressed) 
suggests that that the typical star formation timescales are probably
a substantial fraction of the Hubble time. Note that instantaneously quenched galaxies 
may fade too quickly to be selected as dropout galaxies.
\item The first Spitzer/IRAC detection of $z=8$ galaxies and their red
average $H_{160}-[3.6]\approx0.55$ suggest that luminous early galaxies may 
have substantial $M/L_V\approx0.15$, similar to $z=7$ galaxies. 
The derived stellar mass density then increases gradually with time following 
$\log\rho^*(z)=10.6(\pm0.6) - 4.4(\pm0.7) \log(1+z)$ $[M_\sun~$Mpc$^{-3}]$ over $3<z< 8$.
\end{itemize}

Deeper IRAC data on $z>7$ galaxies are needed to bolster these results. 
More nebular emission line measurements of $z>2$ galaxies would help
to understand the possible contribution to the broadband fluxes.
Additional modeling of the distribution of SFR versus $M^*$ 
is needed to decipher the SFHs of $z\sim7$ galaxies. 
Larger samples would enable a more secure assessment of the
mass density evolution beyond $z=8$.

\acknowledgments
 
We are grateful to all those at NASA, STScI, JPL, SSC who have made Hubble and 
Spitzer the remarkable observatories that they are today. 
IL acknowledges support from NASA through Hubble Fellowship grant 
HF-01209.01-A awarded by the STScI, which is operated by the AURA, Inc., 
for NASA, under contract NAS 5-26555. PO acknowledges support from the Swiss 
National Foundation (SNF). We acknowledge the support of NASA grant 
NAG5-7697 and NASA grant HST-GO-11563. \\

{\it Facilities:} \facility{Spitzer (IRAC)}, \facility{HST (WFC3/IR)}, \facility{Magellan (PANIC)}, \facility{VLT (ISAAC)}

\setlength{\tabcolsep}{0.015in} 

\begin{deluxetable}{crrrrrrrrrr}
\tabletypesize{\scriptsize}
\tablecaption{Summary of photometry and modeling of $z\sim7-8$ dropout galaxies}
\tablewidth{17cm}
\startdata
 \tableline \\
 \multicolumn{11}{c}{SEDs of $z\sim7$ $z_{850}$-dropouts } \\
 \tableline 
\colhead{} & \colhead{$B_{435}$}  & \colhead{$V_{606}$}  & \colhead{$i_{775}$} 
& \colhead{$z_{850}$} & \colhead{$Y$} & \colhead{$J_{125}$} & 
\colhead{$H_{160}$}  & \colhead{$K$}& \colhead{[3.6]} & \colhead{[4.5]}\\
 \tableline 
ERS-2056344288&4.0(7.1)&1.2(5.0)&6.8(8.6)&13.5(9)&40.9(11)&48.6(8)&61.3(10)&---(---)&223.1(51)&134.7(89)\\
ERS-2068244221&-1.6(6.9)&2.4(5.5)&4.6(9.4)&26.4(10)&80.5(11)&78.3(15)&56.4(10)&---(---)&158.8(51)&-5.5(89)\\
ERS-2111644168&6.8(6.2)&1.5(4.7)&7.4(7.5)&4.9(8)&20.1(8)&54.8(10)&43.1(8)&---(---)&142.8(83)&-54.0(89)\\
ERS-2150242362&6.1(8.5)&0.9(7.5)&-3.5(12.5)&14.3(15)&33.4(15)&49.0(10)&68.7(14)&---(---)&154.4(58)&98.9(106)\\
ERS-2150943417&5.2(6.7)&4.7(6.0)&4.4(9.9)&9.2(12)&34.4(16)&87.7(17)&79.7(15)&---(---)&132.4(63)&197.3(120)\\
ERS-2154043286&-0.1(5.0)&-1.4(4.3)&4.8(7.4)&17.7(8)&45.3(9)&45.0(11)&36.5(9)&---(---)&11.5(62)&-13.5(100)\\
ERS-2160041591&-6.0(9.9)&-12.8(7.9)&4.7(13.5)&24.2(16)&68.2(15)&89.8(16)&74.7(14)&---(---)&271.0(87)&293.9(134)\\
ERS-2161941498&-0.8(7.1)&-1.4(5.7)&-3.2(9.8)&8.4(12)&25.7(10)&42.9(10)&41.6(10)&---(---)&64.4(51)&83.2(89)\\
ERS-2202443342&7.2(8.8)&-3.0(7.2)&5.6(10.3)&25.4(11)&42.5(11)&55.2(12)&50.4(11)&---(---)&145.9(51)&111.1(89)\\
ERS-2225241173&1.8(5.6)&-2.0(4.4)&-11.2(7.6)&15.3(8)&37.7(9)&40.6(8)&43.1(8)&---(---)&89.0(51)&128.2(89)\\
ERS-2226543006&-2.5(9.8)&-8.7(7.6)&7.8(12.5)&48.6(14)&113.3(15)&125.1(9)&182.5(13)&---(---)&462.9(94)&276.5(152)\\
ERS-2229344099&-0.6(8.3)&-0.6(7.0)&6.2(10.3)&18.8(11)&49.8(13)&47.0(8)&64.4(11)&---(---)&216.3(51)&110.8(89)\\
ERS-2295342044&-0.2(7.8)&1.4(5.7)&3.7(10.2)&14.5(12)&54.8(13)&70.9(13)&67.1(12)&---(---)&124.7(51)&82.3(89)\\
ERS-2352941047&-3.7(6.0)&4.1(4.6)&4.9(7.6)&8.6(9)&19.7(7)&28.9(6)&31.9(7)&---(---)&34.4(51)&54.4(89)\\
ERS-2354442550&3.7(6.2)&-0.7(4.6)&0.7(7.9)&15.1(9)&46.7(12)&117.4(12)&106.5(11)&---(---)&317.6(53)&21.5(101)\\
 \tableline 
$25<H_{160}<26.5$&-3.2(2.3)&2.8(2.3)&-7.4(4.7)&23.5(5.4)&63.7(7.7)&129.7(16)&128.2(17)&101(12)&262(28)&181(37)\\
$26.5<H_{160}<27.5$&-2.4(0.8)&-0.8(0.6)&2.2(1.8)&14.6(2.3)&49.1(4.3)&64.8(4.1)&57.0(3.4)&24.5(24)&107(16)&83.8(25)\\
$H_{160}>27.5$&1.2(0.8)&0.4(0.5)&-0.1(0.7)&5.6(1.1)&27.2(1.8)&29.6(2.0)&24.5(1.6)&22.9(14)&45.1(9.5)&39.3(17)\\
 \tableline \\
 \multicolumn{11}{c}{SEDs of $z\sim8$ $Y_{098}$-dropouts } \\ 
 \tableline 
\colhead{} & \colhead{$B_{435}$}  & \colhead{$V_{606}$}  & \colhead{$i_{775}$} 
& \colhead{$z_{850}$} & \colhead{$Y_{098}$} & \colhead{$J_{125}$} & 
\colhead{$H_{160}$}  & \colhead{$K$}& \colhead{[3.6]} & \colhead{[4.5]}\\
 \tableline 
ERSy-2354441327&-2.5(6.1)&-2.3(4.7)&-1.4(7.6)&-8.3(9)&1.6(8)&40.7(8)&37.3(7)&---(---)&29.9(52)&61.7(89)\\
ERSy-2376440061&-4.7(8.3)&2.6(5.5)&0.8(10.6)&15.0(11)&6.2(13)&46.5(12)&49.2(13)&---(---)&134.1(51)&143.6(89)\\
ERSy-2251641574&0.4(9.4)&-3.5(6.0)&2.1(9.9)&9.3(11)&5.9(14)&59.3(11)&69.6(13)&---(---)&103.1(51)&38.9(89)\\
 \tableline 
$H_{160}\sim27$&-1.9(4.6)&-1.3(3.1)&0.9(5.4)&6.9(6)&5.0(7)&50.8(6)&55.5(6)&---(---)&95.4(26)&77.4(49) \\
\tableline\\
 \multicolumn{5}{c}{Average colors of $z\sim7$ $z_{850}$-dropouts } & \multicolumn{5}{c}{SFRs and stellar masses of $z\sim7$ $z_{850}$-dropouts }\\ 
\tableline
\colhead{} & \colhead{$H_{160}$} & \colhead{$J_{125}-H_{160}$}  & \colhead{$H_{160}-[3.6]$} &
\colhead{} & \colhead{} & \colhead{} &  \colhead{$SFR_{1500}$} & \colhead{$M^*$} & \colhead{$SSFR$}  & \colhead{} \\
\colhead{}  & \colhead{[AB mag]} & \colhead{[AB mag]}  & \colhead{[AB mag]}   
& \colhead{}  & \colhead{} & \colhead{}  & \colhead{[$M_\sun~$yr$^{-1}$]}  & \colhead{[$M_\sun$]} & \colhead{[yr$^{-1}$]} & \colhead{}  \\
\tableline  
$25<H_{160}<26.6$ & 26.13(0.14) &   0.03(0.06) &   0.89(0.19) &  \vline   & \multicolumn{2}{c}{$\log SFR > 1.0 $} &    $1.17^{+0.05}_{-0.04}$ & $9.92^{+0.06}_{-0.11}$ & -8.75$^{+0.14}_{-0.08}$\\
$26.6<H_{160}<27.5$ & 26.98(0.05) &  -0.09(0.06) &   0.78(0.17)  & \vline  & \multicolumn{2}{c}{$1.0<\log SFR<0.6$}&   $0.85^{+0.02}_{-0.02}$ & $9.55^{+0.08}_{-0.07}$ & -8.70$^{+0.12}_{-0.09}$\\  
$H_{160}>27.5$ & 27.92(0.10) &  -0.20(0.05) &   0.59(0.28)     &  \vline   & \multicolumn{2}{c}{$\log SFR<0.6$} &       $0.39^{+0.05}_{-0.07}$ & $9.08^{+0.12}_{-0.30}$ & -8.70$^{+0.11}_{-0.07}$
\enddata
\tablecomments{
The optical--to--near-IR 
fluxes are measured in $0\farcs4$ diameter apertures. Spitzer/IRAC fluxes
are measured on the confusion-corrected maps in $2\farcs5$ diameter apertures. 
Fluxes are corrected to total assuming point source profiles.
Units are nanoJy for the SEDs and AB magnitudes for the average colors. 
The total sample consists of 15 new sources from the WFC3/ERS sample
(Bouwens et al., in preparation), 9 sources from the NICMOS sample
(UDF-1417,964,GNS-1,2,3,4,5,CDFS-4627,HDFN-1216; Gonzalez et al.
2010), and 12 from the WFC3/UDF sample (UDFz-4471,4257,3955,
3958,3722,4314,3677,3744,4056,3638,3973,3853; Oesch et al. 2010,
Labb\'e et al. 2010). 
The stacked $Y-$band of the $z\sim7$ galaxies is a combination of the $Y_{105}$ and $Y_{098}$ bands. 
SFR$_{1500}$ is the SFR derived from the 1500$\AA$ monochromatic
luminosity using the prescription of \cite{Ma98} and corrected for dust using the 
best-fit $A_V$ (mean $<A_V>=0.13$ mag). 
The stellar masses are based on BC03 $0.2_{ }Z_\sun$ models using exponentially 
declining SFHs with $8<\log\tau<11$ and reddening $A_V<0.3$. Uncertainties are
determined by bootstrapping.} 
\end{deluxetable}

\end{document}